# Design and Analysis of Robust Deep Learning Models for Stock Price Prediction

Jaydip Sen[1*] and Sidra Mehtab[2]
[1]Department of Data Science, Praxis Business School, Kolkata, India.
[2]School of Computing and Analytics, NSHM Knowledge Campus, Kolkata, India
*Corresponding author email: jaydip.sen@acm.org.

**Abstract**

Building predictive models for robust and accurate prediction of stock prices and stock price movement is a challenging research problem to solve. The well-known efficient market hypothesis believes in the impossibility of accurate prediction of future stock prices in an efficient stock market as the stock prices are assumed to be purely stochastic. However, numerous works proposed by researchers have demonstrated that it is possible to predict future stock prices with a high level of precision using sophisticated algorithms, model architectures, and the selection of appropriate variables in the models. This chapter proposes a collection of predictive regression models built on deep learning architecture for robust and precise prediction of the future prices of a stock listed in the diversified sectors in the National Stock Exchange (NSE) of India. The Metastock tool is used to download the historical stock prices over a period of two years (2013- 2014) at 5 minutes intervals. While the records for the first year are used to train the models, the testing is carried out using the remaining records. The design approaches of all the models and their performance results are presented in detail. The models are also compared based on their execution time and accuracy of prediction.

## 1. Introduction

Building predictive models for robust and accurate prediction of stock prices and stock price movement is a very challenging research problem. The well-known efficient market hypothesis precludes any possibility of accurate prediction of future stock prices since it assumes stock prices to be purely stochastic in nature. Numerous works in the finance literature have shown that robust and precise prediction of future stock prices is using sophisticated machine learning and deep learning algorithms, model architectures, and selection of appropriate variables in the models.
Technical analysis of stocks has been a very interesting area of work for the researchers engaged in security and portfolio analysis. Numerous approaches to technical analysis have been proposed in the literature. Most of the algorithms here work on searching and finding some pre-identified patterns and sequences in the time series of stock prices. Prior detection of such patterns can be useful for the investors in the stock market in formulating their investment strategies in the market to maximize their profit. A rich set of such patterns has been identified in the finance literature for studying the behavior of stock price time series.

In this chapter, we propose a collection of forecasting models for predicting the prices of a critical stock of the automobile sector of India. The predictive framework consists of four CNN regression models and six models of regression built on the long-and-short-term memory (LSTM) architecture. Each model has a different architecture, different shapes of the input data, and different hyperparameter values.

The current work has the following three contributions. First, unlike the currently existing works in the literature, which mostly deal with time-series data of daily or weekly stock prices, the models in this work are built and tested on stock price data at a small interval of 5 minutes. Second, our propositions exploit the power of deep learning, and hence, they achieve a very high degree of precision and robustness in their performance. Among all models proposed in this work, the lowest ratio of the root mean square error (RMSE) to the average of the target variable is 0.006967. Finally, the speed of execution of the models is very fast. The fastest model requires 174.78 seconds for the execution of one round on the target hardware platform. It is worth mentioning here that the dataset used for training has 19500 records, while models are tested on 20500 records.

The chapter is organized as follows. Section 2 briefly discusses some related works in the literature. In Section 3, we discuss the method of data acquisition, the methodology followed, and the design details of the ten predictive models proposed by us. Section 4 exhibits the detailed experimental results and their analysis. A comparative study of the performance of the models is also made. In Section 5, we conclude the chapter and identify a few new directions of research.

## 2. Related Work

The literature on systems and methods of stock price forecasting is quite rich. Numerous proposals exist on the mechanisms, approaches, and frameworks for predicting future stock prices and stock price movement patterns. At a broad level, these propositions can be classified into four categories. The proposals of the first category are based on different variants of univariate and multivariate regression models. Some of the notable approaches under this category are - ordinary least square (OLS) regression, multivariate adaptive regression spline (MARS), penalty-based regression, polynomial regression, etc. [2, 13, 16, 37]. These approaches are not, in general, capable of handling the high degree of volatility in the stock price data. Hence, quite often, these models do not yield an acceptable level of accuracy in prediction. Autoregressive integrated moving average (ARIMA) and other approaches of econometrics such as cointegration, vector autoregression (VAR), causality tests, and quantile regression (QR), are some of the methods which fall under the second category of propositions [1, 12, 17, 33, 38, 40-43, 45, 52, 55]. The methods of this category are superior to the simple regression-based methods. However, if the stock price data are too volatile and exhibit strong randomness, the econometric methods also are found to be inadequate, yielding inaccurate forecasting results. The learning-based approach is the salient characteristic of the propositions of the third category. These proposals are based on various algorithms and architectures of machine learning, deep learning, and reinforcement learning [4, 6, 10, 11, 15, 24-30, 34-36, 39, 44, 46-50, 53, 54, 56]. Since the frameworks under this category use complex predictive models working on sophisticated algorithms and architectures, the prediction accuracies of these models are found to be quite accurate in real-world applications. The propositions of the fourth category are broadly based on hybrid models built of machine learning and deep learning algorithms and architectures and also on the relevant inputs of sentiment and news items extracted from the social web [5, 7, 9, 23, 32, 51]. These models are found to yield the most accurate prediction of future stock prices and stock price movement patterns. The information-theoretic approach and the wavelet analysis have also been proposed in stock price prediction [18, 20]. Several portfolio optimization methods have also been presented in some works using forecasted stock returns and risks [3, 8, 19, 21, 22].

In the following, we briefly discuss the salient features of some of the works under each category. We start with the regression-based proposals.

Enke et al. propose a multi-step approach to stock price prediction using a multiple regression model [13]. The proposition is based on a differential-evolution-based fuzzy clustering model and a fuzzy neural network. Ivanovski et al. present a linear regression and correlation study on some important stock prices listed in the Macedonian Stock Exchange [16]. The results of the work indicate a strong relationship between the stock prices and the index values of the stock exchange. Sen and Datta Chaudhuri analyze the trend and the seasonal characteristics of the capital goods sector and the small-cap sector of India using a time series decomposition approach and a linear regression model [37].

Among the econometric approaches, Du proposes an integrated model combining an ARIMA and a backpropagation neural network for predicting the future index values of the Shanghai Stock Exchange [12]. Jarrett and Kyper present an ARIMA-based model for predicting future stock prices [17]. The study conducted by the authors reveals two significant findings: (i) higher accuracy is achieved by models involving fewer parameters, and (ii) the daily return values exhibit a strong autoregressive property. Sen and Datta Chaudhuri different sectors of the Indian stock market using a time series decomposition approach and predict the future stock prices using different types of ARIMA and regression models [38, 40-45]. Zhong and Enke present a gamut of econometric and statistical models, including ARIMA, generalized autoregressive conditional heteroscedasticity (GARCH), smoothing transition autoregressive (STAR), linear and quadratic discriminant analysis [55].

Machine learning and deep learning models have found widespread applications in designing predictive frameworks for stock prices. Baek and Kim propose a framework called ModAugNet, which is built on an LSTM deep learning model [4]. Chou and Nguyen preset a sliding window metaheuristic optimization method for stock price prediction [10]. Gocken et al. propose a hybrid artificial neural network using harmony search and genetic algorithms to analyze the relationship between various technical indicators of stocks and the index of the Turkish stock market [15]. Mehtab and Sen propose a gamut of models designed using machine learning and deep learning algorithms and architectures for accurate prediction of future stock prices and movement patterns [24-30, 46, 47]. The authors present several models which are built on several variants of convolutional neural networks (CNNs) and long-and-short-term memory networks (LSTMs) that yield a very high level of prediction accuracy. Zhang et al. present a multi-layer perceptron for financial data mining that is capable of recommending buy or sell strategies based on forecasted prices of stocks [55].

The hybrid models use relevant information in the social web and exploit the power of machine learning and deep learning architectures and algorithms for making predictions with a high level of accuracy. Among some well-known hybrid models, Bollen et al. present a scheme for computing the mood states of the public from the Twitter feeds and use the mood states information as an input to a nonlinear regression model built on a self-organizing fuzzy neural network [7]. The model is found to have yielded a prediction accuracy of 86%. Mehtab and Sen propose an LSTM-based predictive model with a sentiment analysis module that analyzes the public sentiment on Twitter and produces a highly accurate forecast of future stock prices [23]. Chen et al. present a scheme that collects relevant news articles from the web, converts the text corpus into a word feature set, and feeds the feature set of words into an LSTM regression model to achieve a highly accurate prediction of the future stock prices [9].

The most formidable challenge in designing a robust predictive model with a high level of precision for stock price forecasting is handling the randomness and the volatility exhibited by the time series. The current work utilizes the power of deep learning models in feature extraction and learning while exploiting their architectural diversity in achieving robustness and accuracy in stock price prediction on very granular time series data.

## 3. Methodology

We propose a gamut of predictive models built on deep learning architectures. We train, validate, and then test the models based on the historical stock price records of a well-known stock listed in the NSE, viz. *Century Textiles*. The historical prices of Century Textiles stock from 31st Dec 2012, a Monday to 9th Jan 2015, a Friday, are collected at 5 minutes intervals using the Metastock tool [31]. We carry out the training and validation of the models using the stock price data from 31st Dec 2012 to 30th Dec 2013. The models are tested based on the records for the remaining period, i.e., from 31st Dec 2013, to 9th Jan 2015. For maintaining uniformity in the sequence, we organize the entire dataset as a sequence of daily records arranged on a weekly basis from Monday to Friday. After the dataset is organized suitably, we split the dataset into two parts – the training set and the test set. While the training dataset consists of 19500 records, there are 20500 tuples in the test data. Every record has five attributes – open, high, low, close, and volume. We have not considered any adjusted attribute (i.e., adjusted close, adjusted volume, etc.) in our analysis.

We design ten regression models for stock price forecasting using a deep learning approach. For the univariate models, the objective is to forecast the future values of the variable *open* based on its past values. On the other hand, for the multivariate models, the job is to predict the future values of *open* using the historical values of all the five attributes in the stock data. The models are tested following an approach known as *multi-step prediction using a walk-forward validation* [24]. In this method, we use the training data for constructing the models. The models are then used for predicting the daily *open* values of the stock prices for the coming week. As a week completes, we include the actual stock price records of the week in the training dataset. With this extended training dataset, the open values are forecasted with a forecast horizon of 5 days so that the forecast for the days in the next week is available. This process continues till all the records in the test dataset are processed.

The suitability of CNNs in building predictive models for predicting future stock prices has been demonstrated in our previous work [24]. In the current work, we present a gamut of deep learning models built on CNN and LSTM architectures and illustrate their efficacy and effectiveness in solving the same problem.

CNNs perform two critical functions for extracting rich feature sets from input data. These functions are: (1) convolution and (2) pooling or sub-sampling [14]. A rich set of features is extracted by the convolution operation from the input, while the sub-sampling summarizes the salient features in a given locality in the feature space. The result of the final sub-sampling in a CNN is passed on to possibly multiple dense layers. The fully connected layers learn from the extracted features. The fully connected layers provide the network with the power of prediction.

LSTM is an adapted form of a recurrent neural network (RNN) and can interpret and then forecast sequential data like text and numerical time series data [14]. The networks have the ability to memorize the information on their past states in some designated cells in memory. These memory cells are called *gates*. The information on the past states, which is stored in the memory cells, is aggregated suitably at the forget gates by removing the irrelevant information. The input gates, on the other hand, receive information available to the network at the current timestamp. Using the information available at the input gates and the forget gates, the computation of the predicted values of the target variable is done by the network. The predicted value at each timestamp is made available through the output gate of the network [14].

The deep learning-based models we present in this paper differ in their design, structure, and dataflows. Our proposition includes four models based on the CNN architecture and six models built on the LSTM network architecture. The proposed models are as follows. The models have been named following a convention. The first part of the model's name indicates the model type (CNN or LSTM), the second part of the name indicates the nature of the input data (univariate or multivariate). Finally, the third part is an integer indicating the size of the input data to the model (5 or 10). The ten models are as follows:

(i) CNN_UNIV_5 – a CNN model with an input of univariate open values of stock price records of the last week, (ii) CNN_UNIV_10 – a CNN model with an input of univariate open values of stock price records of the last couple of weeks, (iii) CNN_MULTV_10 – a CNN model with an input of multivariate stock price records consisting of five attributes of the last couple of weeks, where each variable is passed through a separate channel in a CNN, (iv) CNN_MULTH_10 – a CNN model with the last couple of weeks' multivariate input data where each variable is used in a dedicated CNN and then combined in a multi-headed CNN architecture, (v) LSTM_UNIV_5 – an LSTM with univariate open values of the last week as the input, (vi) LSTM_UNIV_10 – an LSTM model with the last couple of weeks' univariate open values as the input, (vii) LSTM_UNIV_ED_10 – an LSTM having an encoding and decoding ability with univariate open values of the last couple of weeks as the input, (viii) LSTM_MULTV_ED_10 – an LSTM based on encoding and decoding of the multivariate stock price data of five attributes of the last couple of weeks as the input, (ix) LSTM_UNIV_CNN_10 – a model with an encoding CNN and a decoding LSTM with univariate open values of the last couple of weeks as the input, and (x) LSTM_UNIV_CONV_10 – a model having a convolutional block for encoding and an LSTM block for decoding and with univariate open values of the last couple of weeks as the input.

We present a brief discussion on the model design. All the hyperparameters (i.e., the number of nodes in a layer, the size of a convolutional, LSTM or pooling layer, etc.) used in all the models are optimized using grid-search. However, we have not discussed the parameter optimization issues in this work.

### 3.1 The CNN_UNIV_5 model

This CNN model is based on a univariate input of open values of the last week's stock price records. The model forecasts the following five values in the sequence as the predicted daily open index for the coming week. The model input has a shape (5, 1) as the five values of the last week's daily open index are used as the input. Since the input data for the model is too small, a solitary convolutional block and a subsequent max-subsampling block are deployed. The convolutional block has a feature space dimension of 16 and the filter (i.e., the kernel) size of 3. The convolutional block enables the model to read each input three times, and for each reading, it extracts 16 features from the input. Hence, the output data shape of the convolutional block is (3,16). The max-pooling layer reduces the dimension of the data by a factor of 1/2. Thus, the max-pooling operation transforms the data shape to (1, 16). The result of the max-pooling layer is transformed into an array structure of one dimension by a flattening operation. This one-dimensional vector is then passed through a dense layer block and fed into the final output layer of the model. The output layer yields the five forecasted *open* values in sequence for the coming week. A batch size of 4 and an epoch number of 20 are used for training the model. The *rectified linear unit* (ReLU) activation function and the Adam optimizer for the gradient descent algorithm are used in all layers except the final output layer. In the output layer of the model, the sigmoid is used as the activation function. The use of the activation function and the optimizer is the same for all the models. The schematic architecture of the model is depicted in **Figure 1**.

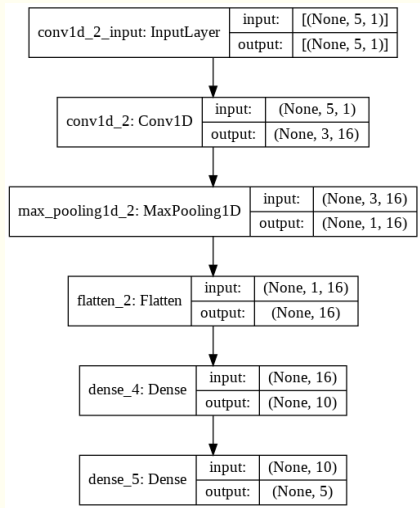

**Figure 1:** The schematic architecture of the model CNN_UNIV_5

We compute the number of trainable parameters in the CNN_UNIV_5 model. As the role of the input layer is to provide the input data to the network, there is no learning involved in the input layer. There is no learning in the pooling layers as all these layers do is calculate the local aggregate features. The flatten layers do not involve any learning as well. Hence, in a CNN model, the trainable parameters are involved only in the convolutional layers and the dense layers. The number of trainable parameters ($n1$) in a one-dimensional convolutional layer is given by (1), where $k$ is the kernel size, and $d$ and $f$ are the sizes of the feature space in the previous layer and the current layer, respectively. Since each element in the feature space has a bias, the term 1 is added in (1)

$$n1 = (k * d + 1) * f \qquad (1)$$

The number of parameters ($n2$) in a dense layer of a CNN is given by (2), in which $p_{current}$ and $p_{previous}$ refer to the node count in the current layer and the previous layer, respectively. The second term on the right-hand side of (2) refers to the *bias* terms for the nodes in the current layer.

$$n2 = (p_{curr} * p_{prev}) + 1 * p_{curr} \qquad (2)$$

The computation of the number of parameters in the CNN_UNIV_5 model is presented in **Table 1**. It is observed that the model involves 289 trainable parameters. The number of parameters in the convolutional layer is 64, while the two dense layers involve 170 and 55 parameters, respectively.

**Table 1:** Computation of the number of params in the model CNN_UNIV_5

| Layer | k | d | f | $p_{prev}$ | $p_{curr}$ | n1 | n2 | #params |
|---|---|---|---|---|---|---|---|---|
| Conv1D (conv1d) | 3 | 1 | 16 | | | 64 | | 64 |
| Dense (dense) | | | | 16 | 10 | | 170 | 170 |
| Dense (dense_1) | | | | 10 | 5 | | 55 | 55 |
| Total #parameters | | | | | | | | 289 |

### 3.2 The CNN_UNIV_10 model

This model is based on a univariate input of the *open* values of the last couple of weeks' stock price data. The model computes the five forecasted daily *open* values in sequence for the coming week. The structure and the data flow for this model are identical to the CNN_UNIV_5 model. However, the input of the model has a shape of (10, 1). We use 70 epochs and 16 batch-size for training the model. **Figure 2** shows the architecture of the model CNN_UNIV_10. The computation of the number of parameters in the model CNN_UNIV_10 is exhibited in **Table 2.**

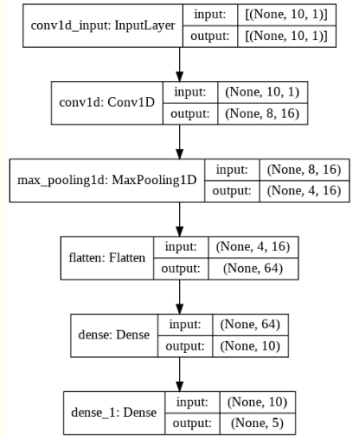

**Figure 2:** The architecture of the model CNN_UNIV_10

**Table 2:** The number of parameters in the model CNN_UNIV_10 model

| Layer | k | d | f | $p_{prev}$ | $p_{curr}$ | n1 | n2 | #params |
|---|---|---|---|---|---|---|---|---|
| Conv1D (conv1d) | 3 | 1 | 16 | | | 64 | | 64 |
| Dense (dense) | | | | 64 | 10 | | 650 | 650 |
| Dense (dense_1) | | | | 10 | 5 | | 55 | 55 |
| Total #parameters | | | | | | | | 769 |

It is evident from **Table 2** that the CNN_UNIV_10 involves 769 trainable parameters. The parameter counts for the convolutional layer, and the two dense layers are 84, 650, and 55, respectively.

### 3.3 The CNN_MULTV_10 model

This CNN model is built on the input of the last two weeks' multivariate stock price records data. The five variables of the stock price time series are used in a CNN in five separate channels. The model uses a couple of convolutional layers, each of size (32,

3). The parameter values of the convolutional blocks indicate that 32 features are extracted from the input data by each convolutional layer using a feature map size of 32 and a filter size of 3. The input to the model has a shape of (10, 5), indicating ten records, each record having five features of the stock price data. After the first convolutional operation, the shape of the data is transformed to (8, 32). The value 32 corresponds to the number of features extracted, while the value 8 is obtained by the formula: $f = (k - n) + 1$, where, $k = 10$, $n = 3$, hence, $f = 8$. Similarly, the output data shape of the second convolutional layer is (6, 32). A max-pooling layer reduces the feature space size by a factor of 1/2 producing an output data shape of (3, 32). The max-pooling block's output is then passed on to a third convolutional layer with a feature map of 16 and a kernel size of 3. The data shape of the output from the third convolutional layer becomes (1, 16) following the same computation rule. Finally, another max-pooling block receives the results of the final convolutional layer. This block does not reduce the feature space since the input data shape to it already (1, 16). Hence, and the output of the final max-pooling layer remains unchanged to (1,16). A flatten operation follows that converts the 16 arrays containing one value to a single array containing 16 values. The output of the flatten operation is passed on to a fully connected block having 100 nodes. Finally, the output block with five nodes computes the predicted daily open index of the coming week. The epochs size and the batch size used in training the model are 70 and 16, respectively. **Figure 3** depicts the CNN_MULTV_10 model. **Table 3** shows the computation of the number of trainable parameters involved in the model.

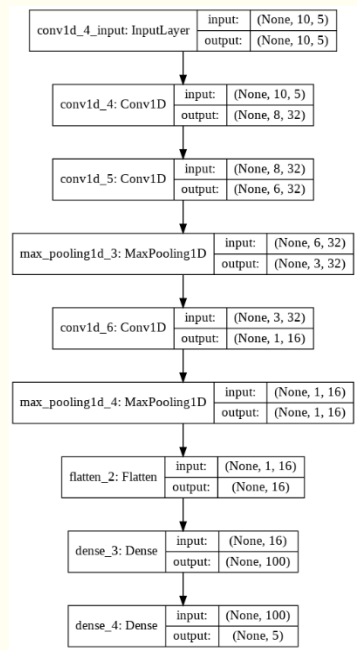

**Figure 3:** The schematic architecture of the model CNN_MULTV_10

From **Table 3**, it is observed that the total number of trainable parameters in the model CNN_MULTV_10 is 7373. The three convolutional layers *conv1d_4*, *conv1d_5*, and *conv1d_6* involve 512, 3014, and 1552 parameters, respectively. It is to be noted that the value of *k* for the first convolutional layer, *conv1d_4*, is multiplied by a factor of five since there are five attributes in the input data for this layer. The two dense layers, *dense_3* and *dense_4* include 1700 and 505 parameters, respectively.

**Table 3:** The number of parameters in the model CNN_MULTV_10

| Layer | k | d | f | $p_{prev}$ | $p_{curr}$ | n1 | n2 | #params |
|---|---|---|---|---|---|---|---|---|
| Conv1D (conv1d_4) | 3*5 | 1 | 32 | | | 512 | | 512 |
| Conv1D (conv1d_5) | 3 | 32 | 32 | | | 3104 | | 3014 |
| Conv1D (conv1d_6) | 3 | 32 | 16 | | | 1552 | | 1552 |
| Dense (dense_3) | | | | 16 | 100 | | 1700 | 1700 |
| Dense (dense_4) | | | | 100 | 5 | | 505 | 505 |
| Total #parameters | | | | | 7373 | | | |

### 3.4 The CNN_MULTH_10 model

This CNN model uses a dedicated CNN block for each of the five input attributes in the stock price data. In other words, for each input variable, a separate CNN is used for feature extrication. We call this a multivariate and multi-headed CNN model. For each sub-CNN model, a couple of convolutional layers were used. The convolutional layers have a feature space dimension of 32 and a filter size (i.e., kernel size) of 3. The convolutional layers are followed by a max-pooling layer. The size of the feature space is reduced by a factor of 1/2 by the max-pooling layer. Following the computation rule discussed under the CNN_MULTV_10 model, the data shape of the output from the max-pooling layer for each sub-CNN model is (3, 32). A flatten operation follows converting the data into a single-dimensional array of size 96 for each input variable. A concatenation operation follows that concatenates the five arrays, each containing 96 values, into a single one-dimensional array of size 96*5 = 480. The output of the concatenation operation is passed successively through two dense layers containing 200 nodes and 100 nodes, respectively. In the end, the output layer having five nodes yields the forecasted five values as the daily open stock prices for the coming week. The epoch number and the batch size used in training the model are 70 and 16, respectively. **Figure 4** shows the structure and data flow of the CNN_MULTH_10 model.

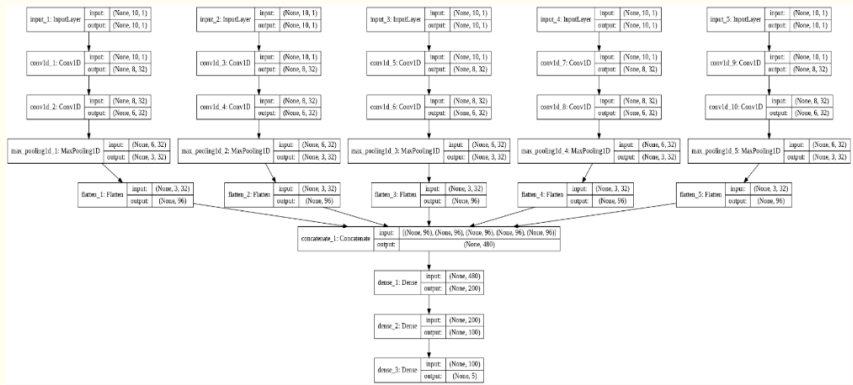

**Figure 4:** The schematic architecture of the model CNN_MULTH_10

**Table 4** presents the necessary calculations for finding the number of parameters in the CNN_MULTH_10 model. Each of the five convolutions layers, *conv1d_1*, *conv1d_3*, *conv1d_5*, *conv1d_7*, and *convid_9*, involves 128 parameters. For each of these layers, k = 3, d = 1 and f = 3, and hence the number of trainable parameters is: (3 * 1 + 1) * 32 = 128. Hence, for the five convolutional layers, the total number of parameters is 128 * 5 = 640. Next, for each of the five convolutional layers, conv1d_2,

conv1d_4, conv1d_6, conv1d_8, and con1d_10, involves 3104. Each layer of this group has $k = 3$, $d = 32$, and $f = 32$. Hence the number of trainable parameters for each layer is: $(3*32 + 1) * 32 = 3104$. Therefore, for the five convolutional layers, the total number of parameters is $3104 * 5 = 15,520$. The dense layers, *dense_1*, *dense_2*, and *dense_3* involve 96200, 20100, and 505 parameters using (2). Hence, the model includes 132,965 parameters.

**Table 4:** The number of parameters in the model CNN_MULTH_10 model

| Layer | k | d | f | $p_{prev}$ | $p_{curr}$ | n1 | n2 | #params |
|---|---|---|---|---|---|---|---|---|
| Conv1D (conv1d_1, conv1d_3, conv1d_5, conv1d_7, conv1d_9) | 3 | 1 | 32 | | | 640 | | 640 |
| Conv1D (conv1d_2, convid_4, conv1d_6, conv1d_8, conv1d_10) | 3 | 32 | 32 | | | 15520 | | 15520 |
| Dense (dense_1) | | | | 480 | 200 | | 96200 | 96200 |
| Dense (dense_2) | | | | 200 | 100 | | 20100 | 20100 |
| Dense (dense_3) | | | | 100 | 5 | | 505 | 505 |
| **Total #parameters** | | | | | 132965 | | | |

### 3.5 The LSTM_UNIV_5 model

This model is based on an input of the univariate information of the open values of the last week's stock price records. The model predicts the future five values in sequence as the daily open index for the coming week. The input has a shape of (5, 1) that indicates that the previous week's daily *open* index values are passed as the input. An LSTM block having 200 nodes receives that data from the input layer. The number of nodes at the LSTM layer is determined using the grid-search. The results of the LSTM block are passed on to a fully connected layer (also known as a dense layer) of 100 nodes. Finally, the output layer containing five nodes receives the output of the dense layer and produces the following five future values of open for the coming week. In training the model, 20 epochs and 16 batch-size are used. **Figure 5** presents the structure and data flow of the model.

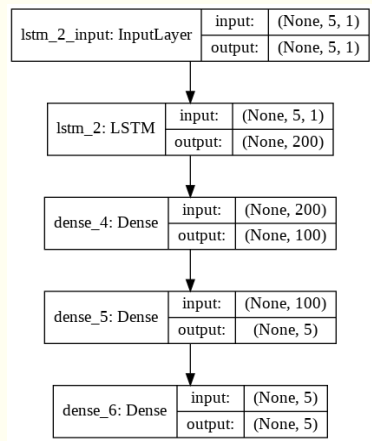

**Figure 5:** The schematic architecture of the model LSTM_UNIV_5

As we did in the case of the CNN models, we now compute the number of parameters involved in the LSTM model. The input layers do not have any parameters, as the role of these layers is to just receive and forward the data. There are four gates in an LSTM network that have the same number of parameters. These four gates are known as (i) forget gate, (ii) input gate, (iii) input modulation gate, and the output gate. The number of parameters ($n1$) in each of the gates in an LSTM network is computed using (3), where $x$ denotes the number of LSTM units, and $y$ is the input dimension (i.e., the number of features in the input data)

$$n1 = (x + y) * x + x \qquad (3)$$

Hence, the total number of parameters in an LSTM layer will be given by 4 * $n1$. The number of parameters ($n2$) in a dense layer of an LSTM network is computed using (4), where $p_{prev}$ and $p_{curr}$ are the number of nodes in the previous layer and the current layer, respectively. The bias parameter of each node in the current layer is represented by the last term on the right-hand side of (4).

$$n2 = (p_{prev} * p_{curr} + p_{curr}) \qquad (4)$$

The computation of the number of parameters associated with the model LSTM_UNIV_5 is depicted in **Table 5**. In Table 5, the number of parameters in the LSTM layer is computed as follows: 4*[(200 + 1) * 200 + 200] = 161,600. The number of parameters in the dense layer, dens_4 is computed as: (200 * 100 + 100) = 20,100. Similarly, the parameters in the dense layers, *dense_5* and *dense_6*, are computed. The total number of parameters in the LSTM_UNIV_5 model is found to be 182, 235.

**Table 5:** The number of parameters in the model LSTM_UNIV_5 model

| Layer | x | y | $p_{prev}$ | $p_{curr}$ | n1 | n2 | #params |
|---|---|---|---|---|---|---|---|
| LSTM (lstm_2) | 200 | 1 | | | 40,400 | | 161600 |
| Dense (dense_4) | | | 200 | 100 | | 20100 | 20100 |
| Dense (dense_5) | | | 100 | 5 | | 505 | 505 |
| Desne (dense_6) | | | 5 | 5 | | 30 | 30 |
| Total #parameters | \multicolumn{7}{c|}{182235} | | | | | | |

### 3.6 The LSTM_UNIV_10 model

LSTM_UNIV_10 model: This univariate model uses the last couple of weeks' open index input and yields the daily forecasted open values for the coming week. The same values of the parameters and hyperparameters of the model LSTM_UNIV_5 are used here. Only, the input data shape is different. The input data shape of this model is (10, 1). **Figure 6** presents the architecture of this model.

**Table 6** presents the computation of the number of parameters involved in the modelLSTM_UNIV_10. Since the number of parameters in the LSTM layers depends only on the number of features in the input data and the node-count in the LSTM layer, and not on the number of input records in one epoch, the model LSTM_UNIV_10 has an identical number of parameters in the LSTM layer as that of the model LSTM_UNIV_5. Since both the models have the same number of dense layers and have the same architecture for those layers, the total number of parameters for both the models are the same.

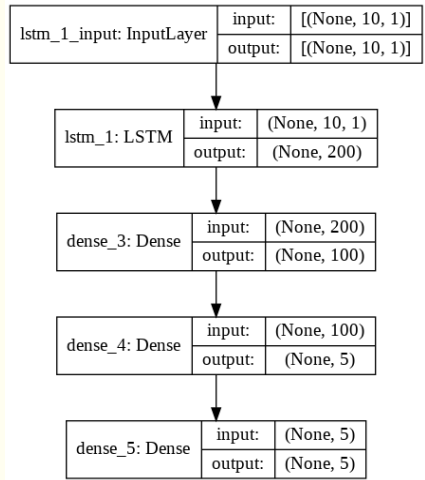

**Figure 6:** The schematic architecture of the model LSTM_UNIV_10

**Table 6:** The number of parameters in the model LSTM_UNIV_10

| Layer | x | y | $p_{prev}$ | $p_{curr}$ | n1 | n2 | #params |
|---|---|---|---|---|---|---|---|
| LSTM (lstm_2) | 200 | 1 | | | 40,400 | | 161600 |
| Dense (dense_4) | | | 200 | 100 | | 20,100 | 20100 |
| Dense (dense_5) | | | 100 | 5 | | 505 | 505 |
| Desne (dense_6) | | | 5 | 5 | | 30 | 30 |
| **Total #parameters** | | | | 182235 | | | |

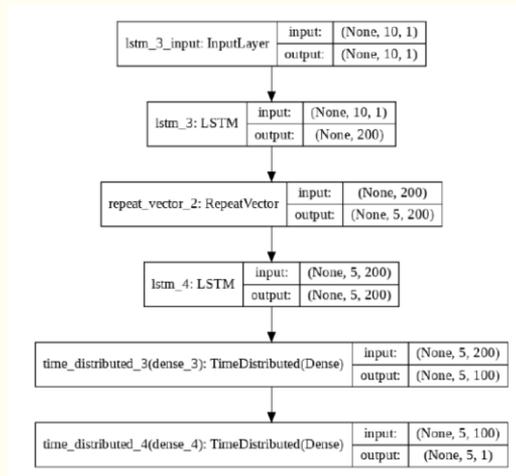

**Figure 7:** The schematic architecture of the model LSTM_UNIV_ED_10

### 3.7 The LSTM_UNIV_ED_10 model

This LSTM model has an encoding and decoding capability and is based on the input of the open values of the stock price records of the last couple of weeks. The model consists of two LSTM blocks. One LSTM block performs the encoding operation, while the other does the decoding. The encoder LSTM block consists of 200 nodes (determined using the grid-search procedure). The input data shape to the encoder LSTM is (10, 1). The encoding layer yields a one-dimensional vector of size 200 – each value corresponding to the feature extracted by a node in the LSTM layer from the ten input values received from the input layer. Corresponding to each timestamp of the output sequence (there are five timestamps for the output sequence for the five forecasted *open* values), the input data features are extracted once. Hence, the data shape from the repeat vector layer's output is (5, 200). It signifies that 200 features are extracted from the input for each of the five timestamps corresponding to the model's output (i.e., forecasted) sequence. The second LSTM block decodes the encoded features using 200 nodes.

**Table 7:** The number of parameters in the model LSTM_UNIV_ED_10

| Layer | x | y | $p_{prev}$ | $p_{curr}$ | n1 | n2 | #params |
|---|---|---|---|---|---|---|---|
| LSTM (lstm_3) | 200 | 1 | | | 40,400 | | 161600 |
| LSTM (lstm_4) | 200 | 200 | | | | 80, 200 | 320800 |
| Dense (time_dist_dense_3) | | | 200 | 100 | | 20,100 | 20100 |
| Dense (time_dist_dense_4) | | | 100 | 1 | | 101 | 101 |
| **Total #parameters** | | | | 502601 | | | |

The decoded result is passed on to a dense layer. The dense layer learns from the decoded values and predicts the future five values of the target variable (i.e., open) for the coming week through five nodes in the output layer. However, the forecasted values are not produced in a single timestamp. The forecasted values for the five days are made in five rounds. The round-wise forecasting is done using a TimeDistributedWrapper function that synchronizes the decoder LSTM block, the fully connected block, and the output layer in every round. The number of epochs and the batch sizes used in training the model are 70 and 16, respectively. **Figure 7** presents the structure and the data flow of the LSTM_UNIV_ED_10 model.

The computation of the number of parameters in the LSTM_UNIV_ED_10 model is shown in **Table 7**. The input layer and the repeat vector layer do not involve any learning, and hence these layers have no parameters. On the other hand, the two LSTM layers, *lstm_3* and *lstm_4*, and the two dense layers, *time_distributed_3*, and *time_distributed_4* involve learning. The number of parameters in the *lstm_3* layer is computed as: 4 * [(200 + 1) * 200 + 200] = 161, 600. The computation of the number of parameters in the *lstm_4* layer is as follows: 4 * [(200 + 200) * 200 + 200] = 320, 800. The computations of the dense layers' parameters are identical to those in the models discussed earlier. The total number of parameters in this model turns out to be 5,02,601.

## 3.8 The LSTM_MULTV_ED_10 model

This model is a multivariate version of LSTM_UNIV_ED_10. It uses the last couple of weeks' stock price records and includes all the five attributes, i.e., *open*, *high*, *low*, *close*, and *volume*. Hence, the input data shape for the model is (10, 5). We use a batch size of 16 while training the model over 20 epochs. **Figure 8** depicts the architecture of the multivariate encoder-decoder LSTM model.

Table 8 shows the number of parameters in the LSTM_MULTV_ED_10 model. The computation of the parameters for this model is exactly similar to that for the model LSTM_UNIV_ED_50 expect for the first LSTM layer. The number of parameters in the first LSTM (i.e., the encoder) layer for this model will be different since the number of parameters is dependent on the count of the features in the input data. The computation of the parameter counts in the encoder LSTM layer, *lstm_1*, of the model is done as follows: 4 * [(200 + 5) * 200 + 200] = 164800. The total number of parameters for the model is found to be 505801.

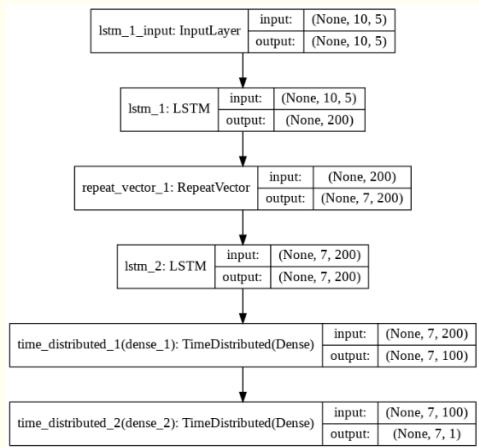

**Figure 8:** The schematic architecture of the model LSTM_MULTV_ED_10

**Table 8:** The number of parameters in the model LSTM_MULTV_ED_10

| Layer | x | y | $p_{prev}$ | $p_{curr}$ | n1 | n2 | #params |
|---|---|---|---|---|---|---|---|
| LSTM (lstm_1) | 200 | 5 | | | 41200 | | 164800 |
| LSTM (lstm_2) | 200 | 200 | | | | 80, 200 | 320800 |
| Dense (time_dist_dense_1) | | | 200 | 100 | | 20,100 | 20100 |
| Dense (time_dist_dense_2) | | | 100 | 1 | | 101 | 101 |
| **Total #parameters** | | | | 505801 | | | |

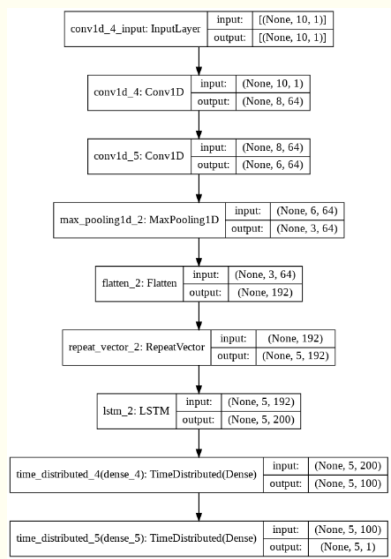

**Figure 9**: The schematic architecture of the model LSTM_UNIV_CNN_10

### 3.9 The LSTM_UNIV_CNN_10 model

This model is a modified version of the LSTM_UNIV_ED_N_10 model. A dedicated CNN block carries out the encoding operation. CNNs are poor in their ability to learn from sequential data. However, we exploit the power of a one-dimensional CNN in extracting important features from time-series data. After the feature extraction is done, the extracted features are provided as the input into an LSTM block. The LSTM block decodes the features and makes robust forecasting of the future values in the sequence. The CNN block consists of a couple of convolutional layers, each of which has a feature map size of 64 and a kernel size of 3. The input data shape is (10, 1) as the model uses univariate data of the target variable of the past couple of weeks. The output shape of the initial convolutional layer is (8, 64). The value of 8 is arrived at using the computation: (10-3+1), while 64 refers to the feature space dimension.

Similarly, the shape of the output of the next convolutional block is (6, 64). A max-subsampling block follows, which contracts the feature-space dimension by 1/2. Hence, the output data shape of the max-pooling layer is (3, 64). The max-pooling layer's output is flattened into an array of single-dimension and size 3*64 = 192. The flattened vector is fed into the decoder LSTM block consisting of 200 nodes. The decoder architecture remains identical to the decoder block of the LSTM_UNIV_ED_10 model. We train the model over 20 epochs, with each epoch using 16 records. The structure and the data flow of the model are shown in **Figure 9**.

**Table 9:** The number of parameters in the model LSTM_UNIV_CNN_10

| Layer | k | d | f | x | y | $p_{prev}$ | $p_{curr}$ | n1 | n2 | #param |
|---|---|---|---|---|---|---|---|---|---|---|
| Conv1D (conv1d_4) | 3 | 1 | 64 | | | | | 256 | | 256 |
| Conv1D (conv1d_5) | 3 | 64 | 64 | | | | | 12352 | | 12352 |
| LSTM (lstm_2) | | | | 200 | 192 | | | 78600 | | 314400 |
| Dense (time_dist_4) | | | | | | 200 | 100 | | 20,100 | 20100 |
| Dense (time_dist_5) | | | | | | 100 | 1 | | 101 | 101 |
| Total #parameters | | | | | | 347209 | | | | |

**Table 9** presents the computation of the number of parameters in the model LSTM_UNIV_CNN_10. The input layer, the max-pooling layer, the flatten operation, and the repeat vector layer do not involve any learning, and hence they have no parameters. The number of parameters in the first convolutional layer is computed as follows: (3 + 1) * 64 = 256. For the second convolutional layer, the number of parameters is computed as: (3 * 64 + 1) * 64 = 12352. The number of parameters for the LSTM layer is computed as follows: 4 * [(200 + 192) * 200 + 200] = 314400. In the case of the first dense layer, the number of parameters is computed as follows: (200 * 100 + 100) = 20100. Finally, the number of parameters in the second dense layer is computed as (100 * 1 + 1) = 101. The total number of parameters in the model is found out to be 347209.

### 3.10 The LSTM_UNIV_CONV_10 model

This model is a modification of the LSTM_UNIV_CNN_10 model. The encoder CNN's convolution operations and the decoding operations of the LSTM sub-module are integrated for every round of the sequence in the output. This encoder-decoder model is also known as the Convolutional-LSTM model [13]. This integrated model reads sequential input data, performs convolution operations on the data without any explicit CNN block, and decodes the extracted features using a dedicated LSTM block. The Keras framework contains a class, ConvLSTM2d, capable of performing two-dimensional convolution operations [13]. The two-dimensional ConvLSTM class is tweaked to enable it to process univariate data of one dimension. The architecture of the model LSTM_UNIV_CONV_10 is represented in **Figure 10**.

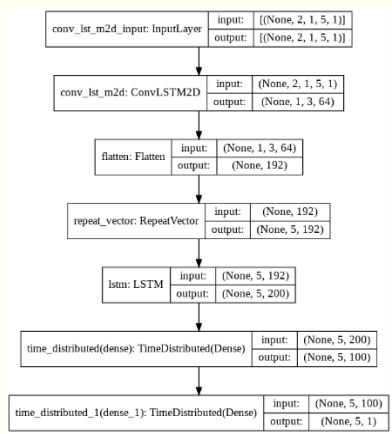

**Figure 10:** The schematic architecture of the model LSTM_UNIV_CONV_10

**Table 10:** Computation of the no. of params in the model LSTM_UNIV_CONV_10

| Layer | k | d | f | x | y | $p_{prev}$ | $p_{curr}$ | n1 | n2 | #param |
|---|---|---|---|---|---|---|---|---|---|---|
| ConvLSTM2D(conv_1st_m2d) | 3 | 1 | 64 | 64 | 1 | | | 12544 | | 50176 |
| LSTM (lstm) | | | | 200 | 192 | | | 78600 | | 314400 |
| Dense (time_dist) | | | | | | 200 | 100 | | 20,100 | 20100 |
| Dense (time_dist_1) | | | | | | 100 | 1 | | 101 | 101 |
| Total #parameters | | | | | | 384777 | | | | |

The computation of the number of parameters for the LSTM_UNIV_CONV_10 model is shown in **Table 10**. While the input layer, the flatten operation, and the repeat vector layer do not involve any learning, the other layers include trainable parameters. The number of parameters in the convolutional LSTM layer (i.e., *conv_1st_m2d*) is computed as follows: $4*x*[k(1+x)+1] = 4*64[3(1+64)+1] = 50176$. The number of parameters in the LSTM layer is computed as follows: $4*[(200+192)*200+100] = 314400$. The number of parameters in the *first time distributed dense layer* is computed as $(200*100+100) = 20100$. The computation for the *final dense layer* is as follows: $(100*1+1) = 101$. The total number of parameters involved in the model, LSTM_UNIV_CONV_10 is 38,4777.

## 4. Performance Results

We present the results on the performance of the ten deep learning models on the dataset we prepared. We also compare the performances of the models. For designing a robust evaluation framework, we execute every model over ten rounds. The average performance of the ten rounds is considered as the overall performance of the model. We use four metrics for evaluation: (i) average RMSE, (ii) the RMSE for different days (i.e., Monday to Friday) of a week, (iii) the time needed for execution of one round, and (iv) the ratio of the RMSE to the response variable's (i.e., *open* value's) mean value. The models are trained on 19500 historical stock records and then tested on 20250 records. The mean value of the response variable, *open*, of the test dataset is 475.70. All experiments are carried on a system with an Intel i7 CPU with a clock frequency in the range 2.60 GHz – 2.56 GHz and 16GB RAM. The time needed to complete one round of execution of each model is recorded in seconds. The models are built using the Python programming language version 3.7.4 and the frameworks TensorFlow 2.3.0 and Keras 2.4.5.

**Table 11** shows the results of the performance of the CNN_UNIV_5 model. The model takes, on average, 174.78 seconds to finish its one cycle of execution. For this model, the ratio of RMSE to the mean *open* values is 0.007288. The ratio of the RMSE to the average of the actual *open* values for day1 through day5 are 0.0062, 0.0066, 0.0073, 0.0078, and 0.0083, respectively. Here, day1 refers to Monday, and day5 is Friday. In all subsequent Tables, we will use the same notations. The RMSE values of the model CNN_UNIV_N_5 plotted on different days in a week are depicted in **Figure 11** as per record no 2 in **Table 11**.

**Table 11:** The RMSE and the execution time of the CNN_UNIV_5 model

| No. | Agg RMSE | Day1 | Day2 | Day3 | Day4 | Day5 | Time (sec) |
|---|---|---|---|---|---|---|---|
| 1 | 4.058 | 4.00 | 3.40 | 3.90 | 4.40 | 4.50 | 173.95 |
| 2 | 3.782 | 3.10 | 3.30 | 3.80 | 4.10 | 4.40 | 176.92 |
| 3 | 3.378 | 2.80 | 3.00 | 3.40 | 3.60 | 3.90 | 172.21 |
| 4 | 3.296 | 2.60 | 3.00 | 3.30 | 3.60 | 3.90 | 173.11 |
| 5 | 3.227 | 2.60 | 3.00 | 3.30 | 3.50 | 3.70 | 174.72 |
| 6 | 3.253 | 2.60 | 3.00 | 3.30 | 3.50 | 3.70 | 183.77 |
| 7 | 3.801 | 3.60 | 3.60 | 3.80 | 3.80 | 4.10 | 172.29 |
| 8 | 3.225 | 2.60 | 2.90 | 3.30 | 3.50 | 3.70 | 171.92 |
| 9 | 3.306 | 2.80 | 3.00 | 3.30 | 3.50 | 3.70 | 174.92 |
| 10 | 3.344 | 2.70 | 3.10 | 3.40 | 3.60 | 3.80 | 174.01 |
| **Mean** | **3.467** | 2.94 | 3.13 | 3.48 | 3.71 | 3.94 | **174.78** |
| RMSE/Mean | 0.007288 | 0.0062 | 0.0066 | 0.0073 | 0.0078 | 0.0083 | |

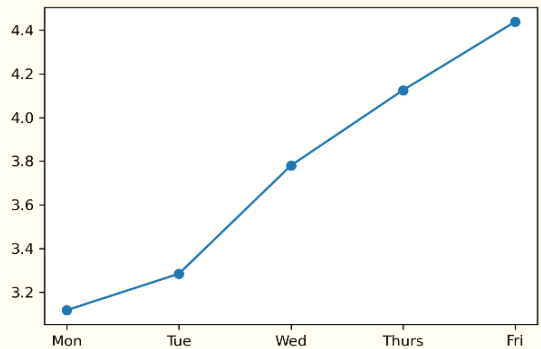

**Figure 11:** RMSE vs. day plot of CNN_UNIV_5 (depicted by tuple#2 in Table 11)

**Table 12** depicts the performance results of the model CNN_UNIV_10. The model needs 185.01 seconds on average for one round. The ratio of the RMSE to the average of the *open* values for the model is 0.006967. The ratios of the RMSE to the average *open* values for day1 through day5 for the model are 0.0056, 0.0067, 0.0070, 0.0075, and 0.0080, respectively. **Figure 12** presents the RMSE values for the results of round 7 in **Table 12**.

**Table 12:** The RMSE and the execution time of the CNN_UNIV_10 model

| No. | Agg RMSE | Day1 | Day2 | Day3 | Day4 | Day5 | Time (sec) |
|---|---|---|---|---|---|---|---|
| 1 | 3.165 | 2.50 | 3.20 | 3.10 | 3.50 | 3.50 | 177.86 |
| 2 | 3.813 | 3.30 | 3.90 | 3.30 | 3.60 | 4.80 | 202.25 |
| 3 | 3.230 | 2.60 | 2.90 | 3.30 | 3.50 | 3.80 | 183.45 |
| 4 | 3.209 | 2.50 | 3.10 | 3.40 | 3.40 | 3.60 | 188.35 |
| 5 | 3.176 | 2.80 | 3.00 | 3.10 | 3.40 | 3.60 | 180.30 |
| 6 | 3.233 | 2.60 | 3.00 | 3.30 | 3.50 | 3.70 | 181.20 |
| 7 | 3.312 | 2.70 | 3.20 | 3.20 | 3.50 | 3.80 | 188.81 |
| 8 | 3.082 | 2.20 | 2.80 | 3.30 | 3.30 | 3.50 | 180.89 |
| 9 | 3.772 | 2.80 | 3.70 | 3.90 | 4.30 | 4.10 | 186.23 |
| 10 | 3.150 | 2.40 | 2.90 | 3.20 | 3.50 | 3.60 | 180.78 |
| **Mean** | **3.3142** | 2.64 | 3.17 | 3.31 | 3.55 | 3.80 | **185.01** |
| **RMSE/Mean** | **0.006967** | 0.0056 | 0.0067 | 0.0070 | 0.0075 | 0.0080 | |

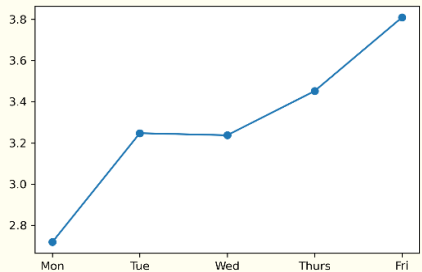

**Figure 12:** RMSE vs. day plot of CNN_UNIV_10 (depicted by tuple#7 in Table 12)

Table 13 depicts the performance results of the model CNN_MULTV_10. One round of execution of the model requires 202.78 seconds. The model yields a value of 0.009420 for the ratio of the RMSE to the average of the *open* values. The ratios of the RMSE values to the mean of the *open* values for day1 through day5 of a week are 0.0085, 0.0089, 0.0095, 0.0100, and 0.0101, respectively. The RMSE values of the model CNN_MULTV_N_10 plotted on different days in a week are depicted in **Figure 13** based on record number 6 of **Table 13**.

**Table 13:** The RMSE and the execution time of the CNN_MULTV_10 model

| No. | Agg RMSE | Day1 | Day2 | Day3 | Day4 | Day5 | Time (sec) |
|---|---|---|---|---|---|---|---|
| 1 | 4.525 | 4.00 | 4.30 | 4.50 | 4.70 | 5.00 | 206.92 |
| 2 | 3.606 | 3.10 | 3.30 | 3.70 | 3.80 | 4.00 | 202.61 |
| 3 | 4.830 | 4.60 | 4.70 | 4.70 | 5.10 | 5.00 | 202.87 |
| 4 | 4.938 | 4.40 | 4.80 | 4.70 | 5.30 | 5.40 | 201.49 |
| 5 | 4.193 | 3.50 | 4.00 | 4.10 | 4.60 | 4.60 | 214.66 |
| 6 | 5.101 | 4.70 | 4.90 | 5.20 | 5.30 | 5.30 | 190.73 |
| 7 | 4.751 | 4.40 | 4.50 | 4.80 | 5.00 | 5.00 | 201.73 |
| 8 | 3.927 | 3.20 | 3.70 | 4.00 | 4.30 | 4.40 | 200.04 |
| 9 | 4.267 | 3.90 | 3.80 | 4.50 | 4.60 | 4.40 | 199.09 |
| 10 | 4.661 | 4.40 | 4.50 | 4.60 | 4.90 | 4.90 | 207.62 |
| **Mean** | **4.4799** | 4.02 | 4.25 | 4.53 | 4.76 | 4.80 | **202.78** |
| **RMSE/Mean** | **0.009420** | 0.0085 | 0.0089 | 0.0095 | 0.0100 | 0.0101 | |

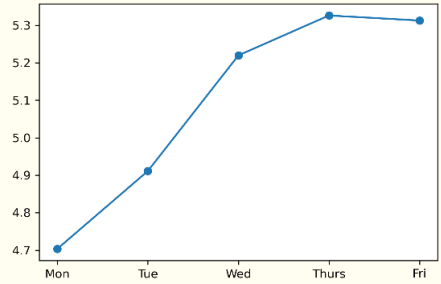

**Figure 13:** RMSE vs. day plot of CNN_MULTV_10 (based on tuple#6 in Table 13)

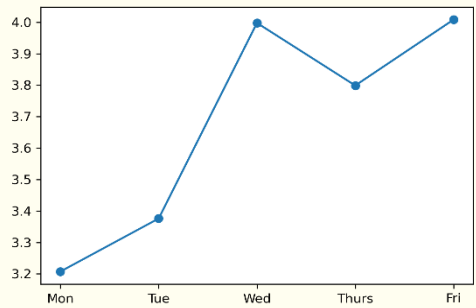

**Figure 14:** RMSE vs. day plot of CNN_MULTH_10 (based on tuple#4 in Table 14)

Table 14 depicts the results of the model CNN_MULTH_10. The model needs, on average, 215.07 seconds to execute its one round. The ratio of the RMSE to the average of the *open* values is 0.008100. The ratios of the RMSE to the average *open* value for day1 to day5 are 0.0076, 0.0075, 0.0082, 0.0084, and 0.0088, respectively. The pattern of variations exhibited by the model daily RMSE is shown in **Figure 14** as per record no 4 in **Table 14**.

**Table 14:** The RMSE and the execution time of the CNN_MULTH_10 model

| No. | Agg RMSE | Day1 | Day2 | Day3 | Day4 | Day5 | Time (sec) |
|---|---|---|---|---|---|---|---|
| 1 | 3.338 | 2.70 | 2.80 | 3.30 | 3.70 | 4.00 | 224.63 |
| 2 | 3.264 | 2.80 | 3.10 | 3.30 | 3.50 | 3.70 | 216.44 |
| 3 | 3.015 | 2.30 | 2.70 | 3.10 | 3.30 | 3.50 | 218.14 |
| 4 | 3.692 | 3.20 | 3.40 | 4.00 | 3.80 | 4.00 | 220.01 |
| 5 | 3.444 | 2.80 | 3.20 | 3.40 | 3.80 | 3.90 | 212.54 |
| 6 | 4.019 | 4.50 | 3.70 | 3.70 | 4.20 | 3.90 | 210.95 |
| 7 | 6.988 | 6.40 | 7.40 | 7.20 | 6.80 | 7.10 | 210.24 |
| 8 | 3.133 | 2.50 | 2.80 | 3.20 | 3.40 | 3.60 | 214.48 |
| 9 | 3.278 | 2.40 | 3.10 | 3.70 | 3.40 | 3.60 | 211.53 |
| 10 | 4.469 | 5.90 | 3.60 | 4.00 | 4.10 | 4.40 | 211.78 |
| **Mean** | **3.864** | 3.55 | 3.58 | 3.89 | 4.00 | 4.17 | **215.07** |
| RMSE/Mean | 0.008100 | 0.0076 | 0.0075 | 0.0082 | 0.0084 | 0.0088 | |

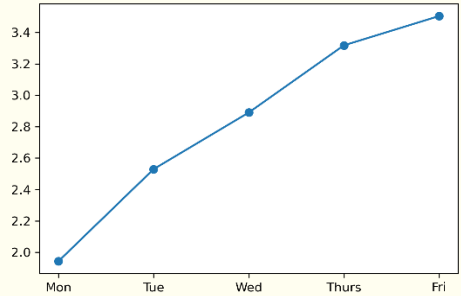

**Figure 15:** RMSE vs. day plot of LSTM_UNIV_5 (depicted by tuple#9 in Table 15)

**Table 15:** The RMSE and the execution time of the LSTM_UNIV_5 model

| No. | Agg RMSE | Day1 | Day2 | Day3 | Day4 | Day5 | Time (sec) |
|---|---|---|---|---|---|---|---|
| 1 | 3.125 | 2.40 | 2.90 | 3.00 | 3.50 | 3.70 | 372.28 |
| 2 | 3.376 | 3.00 | 2.90 | 3.40 | 3.90 | 3.70 | 371.73 |
| 3 | 2.979 | 2.10 | 2.60 | 3.00 | 3.30 | 3.70 | 368.72 |
| 4 | 3.390 | 3.20 | 3.40 | 3.30 | 3.60 | 3.50 | 368.58 |
| 5 | 4.387 | 4.20 | 4.60 | 4.10 | 4.40 | 4.60 | 379.10 |
| 6 | 5.173 | 4.40 | 5.10 | 4.60 | 5.20 | 6.30 | 373.84 |
| 7 | 3.434 | 4.30 | 2.60 | 2.90 | 3.70 | 3.50 | 368.91 |
| 8 | 3.979 | 3.70 | 3.10 | 4.60 | 4.30 | 4.10 | 371.02 |
| 9 | 2.892 | 1.90 | 2.50 | 2.90 | 3.30 | 3.50 | 371.95 |
| 10 | 3.683 | 2.70 | 4.00 | 3.30 | 3.50 | 4.60 | 370.07 |
| **Mean** | **3.6418** | 3.19 | 3.37 | 3.51 | 3.87 | 4.12 | **371.62** |
| **RMSE/Mean** | **0.007770** | 0.0067 | 0.0071 | 0.0074 | 0.0081 | 0.0086 | |

The results of the *LSTM_UNIV_5* model are depicted in **Table 15**. The average time needed to complete one round of the model is 371.62 seconds. The ratio of the RMSE and the average value of the target variable is 0.007770. The RMSE values for day1 to day5 are 0.0067, 0.0071, 0.0074, 0.0081, and 0.0086, respectively. The pattern of variation of the daily RMSE is as per record no 9 in **Table 15** is depicted in **Figure 15**.

**Table 16:** The RMSE and the execution time of the LSTM_UNIV_10 model

| No. | Agg RMSE | Day1 | Day2 | Day3 | Day4 | Day5 | Time (sec) |
|---|---|---|---|---|---|---|---|
| 1 | 3.005 | 2.40 | 2.40 | 2.80 | 3.70 | 3.50 | 547.22 |
| 2 | 3.859 | 3.50 | 3.30 | 3.80 | 3.90 | 4.70 | 554.03 |
| 3 | 4.601 | 4.50 | 4.50 | 4.60 | 4.80 | 4.60 | 550.24 |
| 4 | 3.342 | 2.70 | 4.00 | 3.10 | 3.40 | 3.50 | 555.50 |
| 5 | 4.714 | 4.80 | 4.40 | 4.70 | 4.60 | 5.10 | 563.44 |
| 6 | 3.336 | 2.50 | 3.20 | 3.30 | 3.60 | 3.90 | 553.83 |
| 7 | 3.711 | 3.10 | 4.00 | 4.00 | 3.60 | 3.90 | 559.31 |
| 8 | 2.795 | 1.90 | 2.40 | 2.80 | 3.20 | 3.40 | 552.50 |
| 9 | 3.012 | 1.80 | 2.80 | 2.90 | 3.60 | 3.50 | 551.20 |
| 10 | 2.751 | 1.70 | 2.30 | 3.00 | 3.00 | 3.30 | 557.39 |
| **Mean** | **3.5126** | 2.89 | 3.33 | 3.50 | 3.74 | 3.94 | **554.47** |
| **RMSE/Mean** | **0.007380** | 0.0061 | 0.0070 | 0.0074 | 0.0079 | 0.0083 | |

Table 16 exhibits the results of the model *LSTM_UNIV_10*. The model yields a value of 0.007380 for the ratio of its RMSE to the average *open* values, while one round of its execution needs 554.47 seconds. The RMSE values for day1 to day5 are 0.0061, 0.0070, 0.0074, 0079, and 0.0083 respectively. The RMSE pattern of the model as per record no 10 in **Table 16** is exhibited in **Figure 16**.

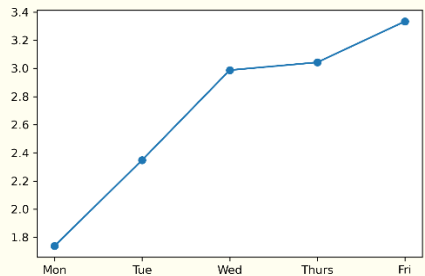

**Figure 16:** RMSE vs. day plot of LSTM_UNIV_10 (depicted by tuple#10 in Table 16)

**Table 17:** The RMSE and the execution time of the LSTM_UNIV_ED_10 model

| No. | Agg RMSE | Day1 | Day2 | Day3 | Day4 | Day5 | Time (sec) |
|---|---|---|---|---|---|---|---|
| 1 | 2.975 | 2.00 | 2.70 | 3.00 | 3.40 | 3.60 | 310.28 |
| 2 | 4.856 | 4.10 | 4.60 | 5.00 | 5.20 | 5.30 | 306.22 |
| 3 | 5.500 | 4.30 | 5.20 | 5.50 | 6.00 | 6.40 | 306.08 |
| 4 | 3.656 | 3.20 | 3.40 | 3.70 | 3.90 | 4.10 | 305.64 |
| 5 | 2.859 | 1.90 | 2.60 | 2.90 | 3.20 | 3.40 | 306.03 |
| 6 | 3.887 | 3.30 | 3.60 | 3.90 | 4.20 | 4.40 | 305.34 |
| 7 | 4.007 | 3.60 | 3.70 | 4.00 | 4.10 | 4.50 | 304.69 |
| 8 | 3.489 | 2.70 | 3.20 | 3.60 | 3.80 | 3.90 | 305.26 |
| 9 | 2.944 | 2.10 | 2.80 | 3.00 | 3.20 | 3.50 | 314.37 |
| 10 | 5.497 | 4.70 | 5.10 | 5.60 | 6.00 | 5.90 | 308.78 |
| **Mean** | **3.971** | 3.19 | 3.69 | 4.02 | 4.30 | 4.50 | **307.27** |
| RMSE/Mean | 0.008350 | 0.0067 | 0.0078 | 0.0085 | 0.0090 | 0.0095 | |

**Table 17** shows that the model LSTM_UNIV_ED_10 needs, on average, 307.27 seconds to execute its one round. The average value of the ratio of the RMSE to the average value of the target variable (i.e., the *open* values) for the model is 0.008350. The daily ratio values for day1 to day 5 of the model are, 0.0067, 0.0078, 0.0085, 0.0090, and 0.0095, respectively. **Figure 17** exhibits the pattern of variation of the daily RMSE as per record no 9 in **Table 17**.

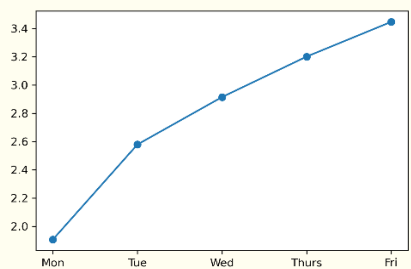

**Figure 17:** RMSE vs. day plot of LSTM_UNIV_ED_10 (as per tuple#5 in Table 17)

Table 18 shows that the model LSTM_MULTV_ED_10, on average, requires 634.34 seconds to complete the execution of its one round. For this model, the ratio of the RMSE to the average value of the target variable (i.e., the *open* values) is 0.010294. The ratios of the daily RMSE to the mean value of *open* for day1 to day5 are, respectively, 0.0094, 0.0099, 0.0102, 0.0107, and 0.0111. **Figure 18** shows the pattern of the daily RMSE values of the model as per record no 10 in **Table 18**.

**Table 18:** The RMSE and the execution time of the LSTM_MULTV_ED_10 model

| No. | Agg RMSE | Day1 | Day2 | Day3 | Day4 | Day5 | Time (sec) |
|---|---|---|---|---|---|---|---|
| 1 | 5.858 | 5.50 | 5.70 | 5.90 | 6.00 | 6.20 | 631.53 |
| 2 | 4.062 | 3.60 | 3.90 | 4.00 | 4.20 | 4.50 | 617.62 |
| 3 | 6.623 | 6.20 | 6.50 | 6.60 | 6.80 | 6.90 | 640.09 |
| 4 | 3.661 | 3.20 | 3.30 | 3.60 | 3.90 | 4.10 | 624.22 |
| 5 | 5.879 | 5.80 | 5.90 | 5.70 | 6.00 | 6.10 | 632.34 |
| 6 | 4.808 | 4.20 | 4.60 | 4.80 | 5.10 | 5.20 | 644.48 |
| 7 | 4.657 | 4.10 | 4.50 | 4.70 | 4.90 | 5.10 | 631.72 |
| 8 | 3.866 | 3.30 | 3.60 | 3.90 | 4.10 | 4.30 | 633.28 |
| 9 | 3.910 | 3.30 | 3.70 | 3.90 | 4.20 | 4.40 | 647.29 |
| 10 | 5.644 | 5.30 | 5.50 | 5.60 | 5.90 | 6.00 | 640.86 |
| **Mean** | **4.897** | 4.50 | 4.72 | 4.87 | 5.11 | 5.28 | **634.34** |
| RMSE/Mean | 0.010294 | 0.0094 | 0.0099 | 0.0102 | 0.0107 | 0.0111 | |

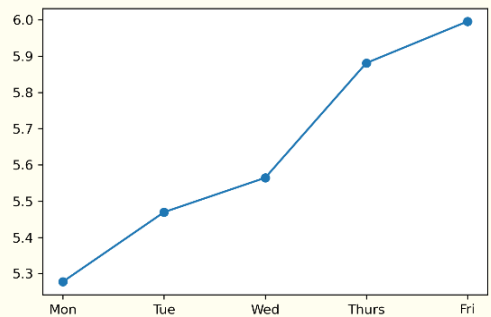

**Figure 18:** RMSE vs. day plot of LSTM_MULTV_ED_10 (as per tuple#10 in Table 18)

**Table 19:** The RMSE and the execution time of the LSTM_UNIV_CNN_10 model

| No. | Agg RMSE | Day1 | Day2 | Day3 | Day4 | Day5 | Time (sec) |
|---|---|---|---|---|---|---|---|
| 1 | 3.832 | 3.30 | 3.50 | 3.90 | 4.10 | 4.30 | 221.18 |
| 2 | 3.256 | 2.50 | 3.00 | 3.30 | 3.60 | 3.80 | 219.74 |
| 3 | 4.308 | 3.80 | 4.00 | 4.40 | 4.60 | 4.60 | 222.59 |
| 4 | 4.081 | 3.30 | 4.00 | 4.10 | 4.30 | 4.50 | 227.95 |
| 5 | 3.325 | 2.60 | 3.00 | 3.30 | 3.60 | 3.90 | 224.46 |
| 6 | 3.870 | 3.20 | 3.70 | 3.90 | 4.10 | 4.10 | 223.40 |
| 7 | 3.688 | 3.10 | 3.40 | 3.80 | 4.00 | 4.10 | 222.89 |
| 8 | 3.851 | 3.20 | 3.60 | 3.80 | 4.20 | 4.40 | 221.87 |
| 9 | 3.710 | 2.60 | 3.40 | 4.00 | 4.00 | 4.40 | 219.74 |
| 10 | 3.736 | 3.30 | 3.70 | 3.70 | 3.90 | 4.10 | 220.96 |
| **Mean** | **3.766** | 3.09 | 3.53 | 3.82 | 4.04 | 4.22 | **222.48** |
| RMSE/Mean | 0.007916 | 0.0065 | 0.0074 | 0.0080 | 0.0085 | 0.0089 | |

**Table 19** depicts that the model LSTM_UNIV_CNN_N_10 requires, on average, 222.48 seconds to finish one round. For this model, the ratio of the RMSE to the average value of the target variable (i.e., the *open* values) is found to be 0.007916. The daily RMSE values for day1 to day5 are, 0.0065, 0.0074, 0.0080, 0.0085, and 0.0089 respectively. **Figure 19** depicts the pattern of variation of the daily RMSE values for this model as per record no 3 in **Table 19**.

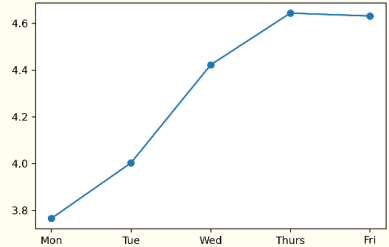

**Figure 19:** RMSE vs. day plot of LSTM_UNIV_CNN_10 (as per tuple#3 in Table 19)

The results of the model LSTM_UNIV_CONV_N_10 are presented in Table 20. The model completes its one round, on average, in 265.97 seconds. The ratio of the RMSE to the average of the *open* values is 0.007490. The daily RMSE for day1 to day5 are 0.0056, 0.0068, 0.0077, 0.0082, and 0.0088, respectively. **Figure 20** shows the patterns of daily RMSE values for this model as per record no 8 in **Table 20**.

**Table 20:** The RMSE and the execution time of the LSTM_UNIV_CONV_10 model

| No. | Agg RMSE | Day1 | Day2 | Day3 | Day4 | Day5 | Time (sec) |
|---|---|---|---|---|---|---|---|
| 1 | 3.971 | 3.00 | 3.60 | 4.00 | 4.40 | 4.60 | 263.84 |
| 2 | 3.103 | 2.40 | 2.80 | 3.20 | 3.40 | 3.60 | 262.06 |
| 3 | 3.236 | 2.30 | 2.90 | 3.30 | 3.60 | 3.80 | 266.47 |
| 4 | 4.347 | 3.10 | 4.00 | 4.60 | 4.70 | 5.00 | 257.43 |
| 5 | 2.860 | 2.20 | 2.50 | 2.80 | 3.20 | 3.40 | 260.05 |
| 6 | 3.525 | 2.50 | 3.60 | 3.50 | 3.80 | 4.00 | 282.27 |
| 7 | 3.163 | 2.30 | 2.80 | 3.20 | 3.50 | 3.80 | 265.26 |
| 8 | 2.870 | 2.00 | 2.60 | 2.90 | 3.20 | 3.50 | 272.18 |
| 9 | 3.504 | 2.20 | 3.10 | 3.70 | 3.70 | 4.40 | 265.46 |
| 10 | 5.053 | 4.70 | 4.40 | 5.20 | 5.30 | 5.60 | 264.66 |
| **Mean** | **3.563** | 2.67 | 3.23 | 3.64 | 3.88 | 4.17 | **265.97** |
| RMSE/Mean | 0.007490 | 0.0056 | 0.0068 | 0.0077 | 0.0082 | 0.0088 | |

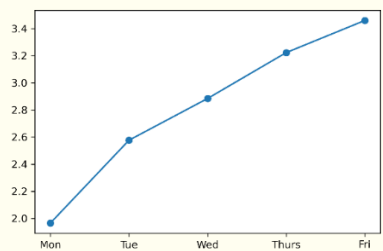

**Figure 20:** RMSE vs. day plot of LSTM_UNIV_CONV_10 (as per tuple#8 in Table 20)

**Table 21:** Comparative analysis of the accuracy and execution speed of the models

| No. | Model | #param | RMSE/Mean | Rank | Exec. Time (s) | Rank |
|---|---|---|---|---|---|---|
| 1 | CNN_UNIV_5 | 289 | 0.007288 | 2 | 174.78 | 1 |
| 2 | CNN_UNIV_10 | 769 | 0.006967 | 1 | 180.01 | 2 |
| 3 | CNN_MULTV_10 | 7373 | 0.009420 | 9 | 202.78 | 3 |
| 4 | CNN_MULTH_10 | 132965 | 0.008100 | 7 | 215.07 | 4 |
| 5 | LSTM_UNIV_5 | 182235 | 0.007770 | 5 | 371.62 | 8 |
| 6 | LSTM_UNIV_10 | 182235 | 0.007380 | 3 | 554.47 | 9 |
| 7 | LSTM_UNIV_ED_10 | 502601 | 0.008350 | 8 | 307.27 | 7 |
| 8 | LSTM_MULTV_ED_10 | 505801 | 0.010294 | 10 | 634.34 | 10 |
| 9 | LSTM_UNIV_CNN_10 | 347209 | 0.007916 | 6 | 222.48 | 5 |
| 10 | LSTM_UNIV_CONV_10 | 384777 | 0.007490 | 4 | 265.97 | 6 |

**Table 21** summarizes the performance of the ten models proposed in this chapter. We evaluate the models on two metrics and then rank the models on the basis of each metric. The two metrics used for the model evaluation are: (i) an accuracy matric computed as the ratio of the RMSE to the mean value of the target variable (i.e., *open* values), and (ii) a speed metric as measured by the time (in seconds) required by the model for execution of its one round. The number of parameters in each model is also presented. It is noted that the CNN_UNIV_5 model is ranked 1 on its execution speed, while it occupies rank 2 on the accuracy parameter. The CNN_UNIV_10 model, on the other hand, is ranked 2 in terms of its speed of execution, while it is the most accurate model. It is also interesting to note that all the CNN models are faster than their LSTM counterparts. However, there is no appreciable difference in their accuracies except for the multivariate encoder-decoder LSTM model, LSTM_MULTV_ED_10.

Another interesting observation is that the multivariate models are found to be inferior to their corresponding univariate models on the basis of the accuracy metric. The multivariate models, CNN_MULTV_10 and LSTM_MULTV_ED_10, are ranked 9 and 10, respectively, under the accuracy metric.

Finally, it is observed that the number of parameters in a model has an effect on its execution time, barring some notable exceptions. For the four CNN models, it is noted that with the increase in the number of parameters, there is a monotone increase in the execution time of the models. For the LSTM models, even though the models, LSTM_UNIV_CNN_10, LSTM_UNIV_CONV_10, and LSTM_UNIV_ED_10, have higher number of parameters than the vanilla LSTM models (i.e., LSTM_UNIV_5 and LSTM_UNIV_10), they are faster in execution. Evidently, the univariate encoder-decoder LSTM models are faster even when they involve a higher number of parameters than the vanilla LSTM models.

## 7. Conclusion

Prediction of future stock prices and price movement patterns is a challenging task if the stock price time series has a large amount of volatility. In this chapter, we presented ten deep learning-based regression models for robust and precise prediction of stock prices. Among the ten models, four of them are built on variants of CNN architectures, while the remaining six are constructed using different LSTM architectures. The historical stock price records are collected using the Metastock tool over a span of two years at five minutes intervals. The models are trained using the records of the first year, and then they are tested on the remaining records. The testing is carried out using an approach known as walk-forward validation, in which, based on the last one- or two-weeks historical stock prices, the predictions of stock prices for the five days of the next week are made. The overall RMSE and the RMSE for each day in a week are computed to evaluate the prediction accuracy of the models. The time needed to complete one round of execution of each model is also noted in order to measure the speed of execution of the models. The results revealed some very interesting observations. First, it is found that while the CNN models are faster, in general, the accuracies of both CNN and LSTM models are comparable. Second, the univariate models are faster and more accurate than their multivariate counterparts. And finally, the number of variables in a model has a significant effect on its speed of execution except for the univariate encoder-decoder LSTM models. As a future scope of work, we will design optimized models based on *generative adversarial networks* (GANs) for exploring the possibility of further improving the performance of the models.